\documentclass[traditabstract]{aa}
\usepackage{txfonts}
\usepackage{natbib}
\bibpunct{(}{)}{;}{a}{}{,}
\usepackage{graphicx}
\usepackage{epstopdf}
\usepackage{multirow} % per celle tabelle che occupano piu' righe
\usepackage{longtable,lscape} % per tabelle lunghe e in landscape

\newcommand{\tab} {Table~}
\newcommand{\fig} {Figure~}

\newcommand{\E} {EDisCS }
\newcommand{\kms} {\rm km \, s^{-1}}
\newcommand{\procedurename}[1]{\texttt{#1}}

\begin{document}
\title{Cl 1103.7$-$1245 at $z=0.96$: the highest redshift galaxy cluster in the EDisCS
survey\thanks{Based on observations collected at the European Southern 
Observatory (ESO) Chile, as part of programme 080.A-0180(A), as well as on programme 166.A-0.162 (the ESO Distant Cluster Survey).}}

\author{Benedetta Vulcani\inst{1,2}\thanks{benedetta.vulcani@oapd.inaf.it}
\and Alfonso Arag{\'o}n-Salamanca\inst{3}%
\and Bianca M. Poggianti\inst{2}%
\and
Bo Milvang-Jensen\inst{4}
\and
Anja von der Linden\inst{5}
\and
Jacopo Fritz\inst{6}
\and 
Pascale Jablonka\inst{7,8}
\and
Olivia Johnson\inst{9}
\and 
Dennis Zaritsky\inst{10}
}

\institute{Astronomical Department, Padova University, vicolo dell'Osservatorio 3, I-35122 Padova, Italy
\and INAF-Astronomical Observatory of Padova, Vicolo dell'Osservatorio 5, I - 35122 Padova, Italy
\and School of Physics and Astronomy, University of Nottingham, Nottingham NG7 2RD, UK
\and Dark Cosmology Centre, Niels Bohr Institute, University of Copenhagen, Juliane Maries Vej 30, 2100 Copenhagen, Denmark
\and Kavli Institute for Particle Astrophysics and Cosmology, Stanford University, Stanford, CA 94309, USA
\and Sterrenkundig Observatorium Vakgroep Fysica en Sterrenkunde Universeit Gent,  Krijgslaan 281, S9  9000 Gent, Belgium
\and Observatoire de Gen\'eve, Laboratoire d'Astrophysique Ecole Polytechnique Federale de Lausanne (EPFL), 1290 Sauverny, Switzerland
\and GEPI, Observatoire de Paris, CNRS UMR 8111, Universit\'e Paris Diderot, 92125 Meudon Cedex, France
\and Royal Observatory,  Blackheath Avenue Greenwich, SE10 8XJ, UK
\and Steward Observatory, University of Arizona, 933 North Cherry Avenue, Tucson, AZ 85721, USA
}
\date{Accepted .... Received ..; in original form ...}

\abstract{We present new spectroscopic observations in a field containing the highest redshift cluster 
of the ESO Distant Cluster  Survey (EDisCS). We measure galaxy redshifts 
and determine the velocity dispersions of the galaxy structures located in this field. Together with the 
main cluster Cl~1103.7$-$1245 ($z=0.9580$; $\sigma_{\rm clus} = 522 \pm 111 \, \kms$) we find a secondary structure
at $z=0.9830$, Cl~1103.7$-$1245c. We then characterize the galaxy properties in both systems, and find that 
they contain very different galaxy populations. The cluster Cl~1103.7$-$1245 hosts a mixture of 
passive elliptical galaxies and star-forming spirals and irregulars.
In the secondary structure Cl~1103.7$-$1245c all galaxies are lower-mass star-forming irregulars and peculiars. 
In addition, we  compare the galaxy populations in the Cl~1103.7$-$1245 $z=0.9580$ cluster  with those in
lower redshift EDisCS clusters with similar velocity dispersions. We find that the properties of the galaxies in 
Cl~1103.7$-$1245 follow the evolutionary trends found at lower redshifts: the number of cluster members increases with time 
in line with the expected growth in cluster mass, and the fraction of passive early-type galaxies increases with time while 
star-forming late types become less dominant. Finally, we find that the mean stellar masses are similar in all clusters, 
suggesting that massive cluster galaxies were already present at $z\sim1$.
 }

\titlerunning{
Cl~1103.7$-$1245 at $z\sim1$: the most distant EDisCS cluster}

\maketitle

\section{Introduction}
Clusters of galaxies, the largest gravitationally-bound systems evolving
from large-scale fluctuations,  provide important constraints on cosmological
models and  are  also  useful laboratories 
to study the effect of  environment on galaxy evolution. In
order to place the strongest constraints on cosmological parameters and explore a broad look-back-time
baseline for evolutionary studies, clusters at redshifts as
high as z$\sim$1 are crucial (e.g., \citealt{aragon93, levine02, Lima04}).

Unfortunately, spectroscopic surveys of sizeable galaxy samples in very distant clusters 
($z\sim1$) are quite rare. At lower redshifts ( $z \sim 0.2$ -- $0.8$), 68 clusters have been studied to-date,
including the spectroscopic surveys of \cite{couch87}, and
the work of the CNOC and MORPHs collaborations \citep{yee96, 
balogh97, dressler99, poggianti99} and,  more recently,  the EDisCS \citep{white05} and ICBS (Oemler et al.\ in preparation)
projects.
In addition  small samples and individual clusters have also been studied (e.g. \citealt{vandokkum00, 
tran03, kelson97, kelson06, bamford05, serote05, moran05}). 

At even higher redshifts, 47 clusters have been studied 
\citep{postman98, postman01, olsen05, gilbank07,muzzin12}, complemented with the analysis of 
individual rich clusters (e.g. \citealt{vandokkum00, 
jorgensen05, jorgensen06, tanaka06, demarco07, tran07,fassbender11}). However, detailed spectroscopic 
studies of clusters at $z \sim 1$
are still relatively few and a clear picture of the galaxy population at this cosmic time has not emerged yet.

For distant cluster studies, 
the ESO Distant Cluster Survey (EDisCS, \citealt{white05}) represents 
an important step toward the understanding of
the star formation histories of cluster galaxies. It
targeted 20 galaxy
clusters, drawn from the Las Campanas Distant Cluster Survey (LCDCS) catalog \citep{gonzalez01},
 with redshifts between 0.4 and 0.8. For these clusters a vast, high-quality, homogeneous
multi-wavelength dataset  has been assembled, making EDisCS arguably the best-studied cluster sample at these redshifts.

Deep spectroscopy with FORS2/VLT was initially obtained for 18 of the fields
\citep{halliday04, milvang08}. 
Spectroscopic targets were selected with the aim 
of producing an unbiased sample of cluster galaxies. 
During the observation of the LCDCS field at $z=0.83$ \citep{gonzalez01},
 a distant cluster at the LCDCS position but at a substantially
  higher redshift ($z = 0.96$) than that inferred from the confirmation
  spectroscopy was discovered %and so built into the spectroscopic targetting
  %strategy 
  (for details, see \citealt{white05}). 
%During the observations of the field of a $z = 0.63$ cluster, a rich $z=0.96$ cluster, {Cl 1103.7$-$1245}, 
%was serendipitously discovered and spectroscopically confirmed.
As  spectroscopic targets were pre-selected using photometric redshifts to minimize field contamination, the \E
spectroscopic sample in this field was biased against $z = 0.96$ galaxies. The discovery of nine galaxies
at this redshift revealed the presence of a moderately rich cluster, with a preliminary spectroscopic velocity
dispersion of  $\sigma^{\rm spec}_{\rm clus} = 600^{+100}_{-150} \, \kms$. 
Given the large number of cluster members found in observations 
biased against them, it seemed likely that the structure was richer still. 

The richness of  Cl 1103.7$-$1245 was confirmed by its clear detection in the weak lensing
analysis of \cite{clowe06}. This study indicates a mass concentration corresponding to a cluster with
a velocity dispersion $\sigma^{\rm lensing}_{\rm clus} = 899^{+129}_{-159}\, \kms$ centred on the brightest cluster galaxy 
with $z = 0.96$. As noted in that paper, however, the lensing analysis might have overestimated
the cluster mass due to the effect of mass belonging to two secondary structures at $z=0.63$ and $z=0.70$. 

Soft diffuse X-ray emission from hot ICM in  Cl 1103.7-1245 was also clearly detected in deep 90ks
XMM-Newton observations (Johnson et al., in preparation).  
The X-ray detection was coincident with both the Brightest Cluster Galaxy (BCG) and the weak-lensing
detection and was of sufficient statistical quality that the X-ray luminosity ($L_X =3.9 ^{+0.5}_{-1.0} \times 10^{43} \,  \mathrm{erg \, s^{-1}}$) 
and temperature ($T= 3.2^{+1.3}_{-0.8} \, \mathrm{keV}$) can be independently constrained by the X-ray spectrum (Johnson et al.,  in preparation). 
\mbox{Cl 1103.7$-$1245} is among the coolest and least luminous $z\sim 1$ clusters to have been detected in X-rays. 

Additional spectroscopy for this cluster was obtained using FORS2 at the VLT, significantly expanding the redshift range of the EDisCS cluster sample. 
In this paper we present these new spectroscopic observations, the data reduction, and a detailed analysis of the galaxies in the field of Cl~1103.7$-$1245.  In addition we compare
this cluster's galaxy properties with those of the galaxy populations of other EDisCS clusters with similar velocity dispersions and different redshifts.
 
Throughout this paper, we assume $H_{0}=70 \, \rm km \, s^{-1} \,
Mpc^{-1}$, $\Omega_{\rm m}=0.30$, $\Omega_{\Lambda} =0.70$.
All magnitudes are in the Vega system.

\section{The data}
\subsection{Target selection and observations}
The target selection strategy, mask design procedure and observations are similar to those adopted 
for the EDisCS spectroscopy, described in detail in
\cite{halliday04} and \cite{milvang08}. 
The only difference is that it targeted  $z=0.96$, unlike the previous target selection that was focused on $z=0.70$.

 The target selection was based on the available
VLT/FORS2 optical photometry \citep{white05} and the
NTT/SOFI NIR photometry (Arag{\'o}n-Salamanca et al., in preparation).
The optical data cover $6.5' \times 6.5'$ and are well-matched to
the FORS2 spectrograph field-of-view.
The NIR data cover a somewhat smaller region of
 $4.2' \times 5.4'$.
The photometry was used as input to a modified version of the photometric
redshifts code \procedurename{hyperz}
\citep{bolzonella00}.
 The aim of the target selection strategy was to keep all galaxies at
the cluster redshift (brighter than $I=23$),
while removing objects that were almost certainly not
galaxies at the cluster redshift.
The selection criteria are explained in \cite{halliday04}. 

Targets already observed in previous runs and  with measured  redshifts  (see \citealt{milvang08})
were preferentially not repeated, when an alternative was found.

 \cite{poirier04} developed a programme to design the spectroscopic
slit masks for \E (called ``MXU masks'' after the Mask eXchange Unit
in the FORS2 spectrograph).
A fuller description of how the programme works
is found in \cite{halliday04}. 
Spectroscopic observations were completed using  
the FORS2 spectrograph\footnote{%
\texttt{http://www.eso.org/instruments/fors}}
(cf. \citealt{appenzeller98}) on the VLT,
during one observing run in March 2008,
comprised of 3 half-nights.
A high-efficiency grism was used 
(grism 600RI+19, $\lambda_\mathrm{central} = 6780\,${\AA},
resolution FWHM $\approx$ 6$\,${\AA}).
The exposure time was 1800 seconds per frame, for a total of 8 frames per mask.\footnote{Only one frame of one mask had an exposure time of  2400 seconds.}
A total of 3 masks were observed, for a total of 94 slits (33 slits in mask 1 and 2 and 28 slits in mask 3).
 
 \subsection{Data reduction}\label{sec:reduc}
 The reduction was performed using an 
``improved sky subtraction'', whose properties and advantages have been largely described and discussed in \cite{milvang08}.
As presented in \cite{milvang03tesi,milvang03,milvang08},
the traditional sky subtraction did not work well for
spectra produced by tilted slits:
strong, systematic residuals were evident where skylines have been subtracted.
As a consequence, an improved method for the sky subtraction was implemented.
The method is described in detail in \citet{kelson03}.
 The key point of this method
is that the sky subtraction is performed prior to
any rebinning/interpolation of the data, hence it results 
in smaller noise than the traditional subtraction.
Briefly, 
starting from the combined but uninterpolated frames,
the data are flat-fielded; then the sky is fitted and subtracted %(as described below)
and the spatial curvature is removed by means of an interpolation in $y$.
The individual spectra are cut-out, the 2D wavelength calibration is applied by means 
of an interpolation in $x$, resulting in rectified 2D spectra
(i.e.\ pixelised in $(x_\mathrm{r},y_\mathrm{t})$) that are sky-subtracted, and with
almost no systematic residuals where the skylines have been subtracted.
Finally 1D spectra are extracted. 

Around the central wavelength, our 1D spectra have a median S/N=4.6 \AA{}  per pixel in the continuum.
\subsection{Galaxy redshifts and equivalent width determinations}\label{sec:galaxy_redshifts}
Galaxy redshifts were  measured from the reduced 1D spectra, using emission lines
where possible, as done in  \cite{poggianti06}. 
We used the [O{\sc ii}]$_{\lambda 3727}$ line, 
the [O{\sc iii}]$_{\lambda 5007}$ line, the H$\alpha_{\lambda 6563}$ line, or
the most prominent absorption lines, e.g.\ Calcium K and H lines at
3934$\,${\AA} and 3968$\,${\AA}. 
The redshifts were manually assigned a quality flag.
The vast majority % ($\sim$95\%)
of the measured redshifts (164/214) are of the highest
quality, and these redshifts are assigned a 0 value in the $z_{quality}$ column in the 
catalogue. 
Secure redshifts but with larger uncertainties are assigned a $flag=1$ (5/214),
and doubtful redshifts are assigned a  $flag=2$ (9/214).
For a fraction of objects
(36/214),
no redshift could be determined,
and these redshifts are listed as 9.9999 in our data tables (flag $z_{quality}= 3$).
 The fraction of galaxies for which redshifts 
could not be obtained is somewhat higher than for the rest of the EDisCS 
clusters \citep{milvang08}. This is due to the fact that this 
cluster is at significantly higher redshift and, although the magnitude 
limit in the target selection is the same, the target galaxies are, on 
average, significantly fainter than for the rest of EDisCS. In other 
words, the magnitude distribution is skewed towards fainter magnitudes, 
explaining the lower success rate.
For the objects targeted as possible cluster members in the 3  long masks,
the statistics are as follows:
10 stars, 204 galaxies and 36 without a determined redshift.
%The quality distribution (i.e. $z_{quality}= 0, 1, 2,3 $) of galaxy redshifts 
%is 76.7\%, 2.3\%, 4.2\%, and 16.8\% respectively.

In our sample, nine objects were %observed more than once, being
already present in \cite{milvang08}. Two of them are stars (which were observed and repeated 
for calibration purposes), 
for one object the \cite{milvang08} redshift estimate was more reliable. In the other cases,
 the difference between the two redshift estimates
was always less than 0.0005. %We can compare this typical redshift error to the %adopt as typical error the 
%error computed by 
\cite{milvang08}, who adopted the same
 instruments, data reduction and procedure % used to compute redshifts
%is exactly the same.
%They 
give an error that corresponds   
to  $ \sim 0.0002$, at any redshift, 
and this is compatible with the one computed  as the difference between the two redshift estimates.

In addition, when possible, we also measured
the equivalent width of [O{\sc ii}] and H$_\delta$.
 Note that for star-forming galaxies the H$_\delta$ absorption line may be 
affected by emission in-filling. We have not attempted to correct for this 
effect. 

From now on, we use both the spectra reduced in the current work %results of my data reduction 
and  those already obtained by \cite{milvang08} for the same field.
 
\subsection{Spectroscopic completeness weights}\label{sec:cl_weight}
During the EDisCS spectroscopic runs, slits were assigned to galaxies giving preference, whenever possible, to the brightest
targets. Therefore, it is
necessary to quantify how the completeness of the spectroscopic samples varies as a function of galaxy apparent magnitude.
This was done comparing the number of objects in the spectroscopic catalogue with
the number in the parent photometric catalogue in bins of $I$ magnitude. 
The parent catalogue included all
entries in the EDisCS photometric catalogue that were retained as targets for spectroscopy (see \citealt{halliday04, milvang08}). 
Since here  we compute spectroscopic completeness also for galaxies observed in \cite{milvang08}, we use as parent catalogue the sum of the parent catalogue
used for the observations presented in \cite{milvang08} plus the parent catalogue adopted for the current observations. 
The ratio of the number of objects in the spectroscopic catalogue to
the number in the parent photometric catalogue yielded a weight as a function of galaxy apparent magnitude.
%This is independent on redshift, hence it is valid also for galaxies at lower {\it z}.  
The mean completeness for galaxies with $I<23$ is $\sim 37\%$.

Geometrical effects, due to possible variations in the sampling as a function of the
cluster-centric radius, can in principle affect a spectroscopic sample of a cluster. However, as discussed in \cite{poggianti06}, geometrical effects are  expected to be small when several masks of the same cluster, always centered on the cluster center, are taken, as it is the case for EDisCS. Hence  we do not have to compute geometrical completeness weights.

\subsection{Morphologies}\label{sec:mor}
The
morphological classification of galaxies have been drawn 
 from \cite{desai07} and it is 
 based on the visual classiÞcation of HST/ACS F814W images sampling the rest-frame
$\sim 4300-5500$ \AA{} range.
 However, since 11/30 %a non negligible fraction 
of our cluster and secondary structure members had
no morphology in \cite{desai07}, being fainter than the magnitude limit adopted in \cite{desai07},
 $ACS-HST$ images were inspected again\footnote{This has been done by three independent classifiers, 
Benedetta Vulcani, Bianca M. Poggianti and Alfonso Arag\'on-Salamanca.} % They already classified \E galaxies.} 
and a new morphology was assigned to the previously unclassified objects.
It was very hard to assign a precise Hubble morphology for them. They are all peculiar, disturbed galaxies, possibly  
undergoing mergers. 

The new morphological classification follows the one proposed by  \cite{desai07}. %, with the exception that peculiar galaxies are assigned a new value. % equal to 20.
The classification is as follows: Star$= 7$, nonstellar but too compact to see structure$=-6$, ellipticals $=-5$,
S0$= 2$, Sa$=1$, Sb$=3$, Sc$=5$, Sd$=7$, Sm$=9$, Irr/Pec$=11$,  no HST data corresponding to ground-based object$=111$. %, unclassifiable$=66$.

\subsection{Determining the galaxy stellar masses}\label{sec:cl_mass}
In previous EDisCS papers (see, e.g, \citealt{morph}),
stellar masses were computed following the method proposed by  \cite{bj01}. 
However, the relation they proposed may not be valid at $z\sim 1$. This is because the spectro-photometric models
they use are %based on stellar population synthesis models calibrated at 
for $z\sim 0$, and assume that galaxies have been forming stars for the last $\sim 13 \, Gyr$. 
In contrast, galaxies change their mass-to-light ratio over time, hence the assumptions that are valid at $z \sim 0$ might not be valid at high-{\it z} and the relation between mass-to-light ratio and colors could change going to higher redshifts.
As a consequence, in this paper we determine
stellar masses %were determined
by fitting the observed photometry with a spectrophotometric model. We then compare the two approaches to
test whether \cite{bj01} method  is still valid at higher-z.

The model used in this paper follows the approach described in \cite{hatz08, hatz09} and was purposely modified to 
run with EDisCS data. It exploits the photometric EDisCS $VRIJK$ bands \citep{white05}.

The aim of the GAlaxies S{\footnotesize ED} FITting code (GASFIT) is to calculate the total stellar mass of galaxies from their observed photometry.
The code uses a combination of theoretical spectra of Simple Stellar Population (SSP) models to reproduce the observed broad-band spectral energy distribution of a galaxy. 

The Padova evolutionary tracks
\citep{bertelli94} and a standard \cite{salpeter55} IMF, with masses in the range $0.15-120 \,  M_\odot$ are used.
For details on the set of SSP see \cite{fritz07, fritz11} and references therein.
The final set of SSPs is composed of 13 spectra
referring to stellar ages ranging from $3\times 10^6$ to $14 \times 10^9$ years.
The set 
has a common metallicity which can be chosen by the user. All the spectra are weighted with a suitable value of the stellar mass, then they are summed. The effect of dust attenuation is taken into account by applying an extinction law, which is simulating dust distributed in a uniform slab sitting in front of the stars; a common value is adopted for all SSPs, i.e. no selective extinction is considered. The Galactic extinction curve is generally adopted, but the code allows for other options as well. The amount of extinction, which is measured through the $E(B-V)$ value, is one of the parameters of the code.

The value of the stellar mass that is used to weight the spectra of each age, is computed by assuming that the star formation history (SFH), that is the star formation rate as a function of time, of a galaxy is well represented by  
a  ``delayed--exponential'' law  \citep{sandage86, gavazzi02}.  

The final model spectrum is hence computed. 
Once the metallicity of the stars and the extinction law are fixed, only two free parameters are used: $E(B-V)$ and $\tau$, which is expressed in terms of the age of the galaxy. To find the combination of these two parameters that minimizes the difference between the observed photometry and the model,  models are calculated for various combinations of $E(B-V)$ and $\tau$, where the first parameter can assume values ranging from 0 to 0.2\footnote{We note that even extending the E(B-V) 
range  up to 1 the results do not change significantly.}
 and the second can range from 0.02 to 1. 
 The best-fit model is chosen as the one yielding the lowest $\chi^2$ value.  This procedure is repeated for each object of the input catalogue.
The total stellar mass values provided by the code include the mass in stars plus stellar remnants. 

Stellar masses have been computed for all galaxies with a reliable redshift.
Since the model adopts a \cite{salpeter55} IMF  in the mass range $0.15-120 \, M_{\odot}$, to convert to  a \cite{kr01} IMF we multiplied the model masses by a factor $1/1.33$.
Given the mass range of the galaxy sample, we used models with solar metallicity. 

We compare the results coming from these models to the masses computed using \cite{bj01} for cluster members  and found that
both methods give compatible mass estimates (rms $\sim 0.1$ dex), 
suggesting that the \cite{bj01} relation   and detailed SED-fitting methods 
give similar results
%works well  
even at $z \sim 0.95$.

To compute the mass completeness limit  
we consider the main structure, at $z\sim0.96$, and determine the value of the
mass of a galaxy with an absolute B magnitude corresponding to $I=23$,
and a color $(B-V) \sim 0.85$, which is the reddest color of galaxies
in this cluster. In this way, the limit for a \cite{kr01} IMF %based on photo-z 
is $M_{\ast}\geq  10^{10.7} M_{\odot}$.

\subsection{The spectroscopic catalogue}
\label{sec:catalogues}
All the previously-available EDisCS spectroscopic information is 
described in \cite{halliday04} and \cite{milvang08}.
The catalogue presented here 
contains all the spectroscopy now available for this cluster field, including our new data and 
the data presented in \cite{milvang08} for completeness. 

The format of the tables is illustrated in \tab\ref{tab:sample_spectable}, 
while the whole catalogue is published electronically at the CDS\@.

Column~1 gives the object name  {\it ID}. %Here I give the new ID 
(IDs starting with EDCS\textbf{N}J), usually adopted to indicate
EDisCS galaxies. 
These IDs are derived from the EDisCS photometric catalogue, 
matching the coordinates (RA and DEC).

Column~2 gives $I_{\rm auto}$, the total $I$-band magnitude (not corrected for Galactic extinction).
This magnitude comes from the catalogues published in \cite{white05}.
 
Column~3 gives the redshift {\it z} as determined in \S\ref{sec:galaxy_redshifts}.
The redshifts are always given with 4 decimal places.
A value of 0.0000 denotes a star, and 9.9999 denotes that no redshift
could be determined.

Column~4 gives the redshift quality $z_{\rm quality}$. 0 means that
redshift measurement is robust. Secure redshifts but with larger uncertainties are assigned a flag=1,
doubtful redshifts are assigned a flag=2, no redshift estimates are assigned a flag=3.

Column~5 gives the membership flag.
It is
  1  for members of the main cluster, 1c, 1d, 1e 
     for members of  secondary structures,
   0    for field galaxies, and
``--'' for stars and objects without a determined redshift.
Membership is defined as being within $\pm 3\ \sigma_{\rm cl}$ from $z_{\rm cl}$.

Column~6 gives $EW_{\rm OII}$, the rest-frame equivalent width, in \AA, of the [O{\sc ii}]$_{\lambda 3727}$ line, 
adopting the convention that EWs are negative when in emission. ``--'' indicates the measurement was not possible
 due to low S/N.

Column~7 gives  $EW_{{\rm H}_{\delta}}$, the rest-frame equivalent width, in \AA, of the H$_{\delta}$  line, 
adopting the convention that EWs are negative when in emission.
``--'' indicates the measurement was not possible  due to low S/N.

Column~8 gives the mass estimates, computed as presented in \S\ref{sec:cl_mass} and converted to a \cite{kr01}
IMF.

Column~9 gives the spectroscopic magnitude completeness weights, computed as presented in \S\ref{sec:cl_weight}.

Column~10 gives the morphology of galaxies. %, when available % (see \S\ref{sec:mor}).
%{\bf In the catalogues, do I call pec/Irr =11?}

Column~11 gives some relevant comments.

\begin{table*}[!h]
\caption{Illustration of the format of the spectroscopic catalogues. \label{tab:sample_spectable}}
\begin{center}
\begin{tabular}{lccccrrrrrc}
\hline
\hline 
ID new &  $ I_{\rm auto}$&     z  & $ z_{\rm qual}$ &memb & $EW_{\rm OII}$  &$EW_{{\rm H}_{\delta}}$ & $ \log M_\ast / M_\odot$ & W &    morph & comm\\ 
\hline
 ...   &...  &...           &...  &... &...  & ... &... &   ...  &... \\                                         
EDCSNJ1103444$-$1245153 &21.534   &0.9640    &0           &1  &0.0   & 2.5   &11.243  &4.71     &$-5$     &...  \\                                      
EDCSNJ1103447$-$1245597 &22.213   &0.9588   &0           &1  &$-20.8$  &4.0   &10.404 &2.58    &3      &...\\                                   
EDCSNJ1103417$-$1245150 &22.188   &0.9885   &0           &1c &$-105.5$ &$-8.0$  &10.087   &3.75  &11   &... \\                                         
EDCSNJ1103437$-$1244540 &22.168   &0.9559   &0        &1  &$-17.8$  &---   &10.538   &3.75      &5     &... \\          
  EDCSNJ1103462$-$1245556 &22.289   &0.7212  &0           &0  &$-66.95$ &---  & 9.748 &   2.58 &    1  &... \\                                         
 ...   &...  &...           &...  &... &...  & ... &... &   ...  &... \\                                         
\hline
\hline
\end{tabular}
\end{center}
\vspace*{0.1ex}
{\footnotesize {\it Notes} -- 
This example table contains entries  to illustrate all relevant features of the tables published electronically at the CDS. This table contains both the new redshifts from data presented in this work (56\% of the redshifts) and all the redshifts taken in this field from the previous EDisCS spectroscopy \citep{milvang08} (44\% of the redshifts). }
\end{table*}

\section{Results}
\subsection{Galaxy positions, redshifts and cluster velocity dispersions}\label{sec:zhist}
In this section, we  study  the distribution of  galaxies in the field, investigating whether they are grouped in structures. We study both their spatial and redshift distributions.  
\begin{figure*}[!t]
\centering
\includegraphics[scale=0.7]{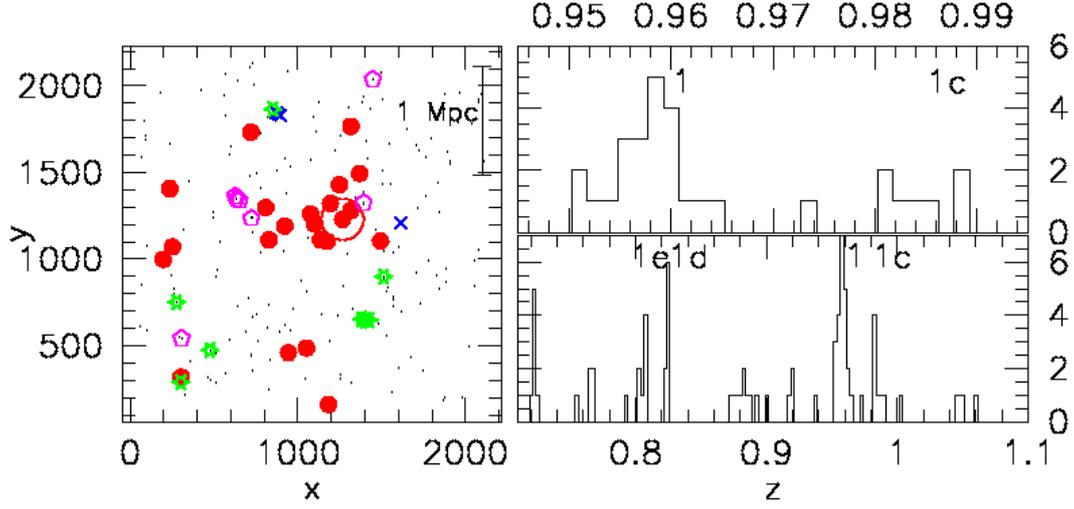}
\caption{ Left panel: spatial distribution of galaxies in the whole field. Pixel positions for galaxies are given. The units on the axes are pixels $= 0.2^{\prime \prime}$. North is up and east is to the left .The structures identified are indicated with the different colors. Black points: all galaxies  in the spectroscopic catalog. Red dots: main cluster, ``1''. Empty magenta pentagons: ``1c''. Green skeletal stars: ``1d''.  Blue crosses: ``1e''. 
The size of 1 Mpc at $z=0.95$ is also shown. The BCG of the main cluster is also indicated with an empty red circle. Lower right panel: redshift histogram of the whole field. The position of the structures identified in the field is also indicated. Upper right panel: redshift histogram of the structures 1 and 1c, which are the structures analyzed in the paper.
\label{fig:histoall}}
\end{figure*}

The lower right panel of \fig\ref{fig:histoall} shows the redshift histogram obtained from the 
spectroscopic observations of the field. Here all galaxies are used,  
both those for which we carried out the data reduction  and those already analyzed by \cite{milvang08}.
In the redshift distribution
four structures can be detected, called 1
(the main peak 
at the cluster redshift), 1c, 1d and 1e,  
characterized in  \tab\ref{tab:structures}.  
\begin{table}[!h]
\caption{ List of the structures identified, with redshift,  and number of members. \label{tab:structures}}
\begin{center}
\begin{tabular}{lcc}
\hline
\hline 
{ name} & { z} &  ${ N_{\rm memb}}$  \\
\hline
1 (cluster) 	& 0.958  & 22 (22)\\
1d		& 0.825  & 8 (6)\\
1c (secondary structure)	& 0.983	  & 7 (6)\\	
1e		& 0.807 &   4 (2) \\
\hline
\hline
\end{tabular}
\end{center}
\vspace*{0.1ex}
{\footnotesize {\it Notes} --  The number of members is within $\pm 3\sigma$ from the structure's redshift. In parenthesis the number of objects with a reliable redshift ($z_{\rm quality}=0$) is given. The structures are labelled in richness order, with the richest (as determined from the number of spectroscopic members) first. }
\end{table}

 The upper right panel of the \fig\ref{fig:histoall} shows a zoom of the redshift histograms
of the structures 1 and 1c that will be analyzed carefully in the following.  %The structures 1 and  1c will be analyzed 
The other structures, 1d  and 1e, will be discarded from our galaxy analysis since they are outside the targeted redshift range  
and the galaxies observed are therefore a biassed subsample of the total population. 

To compute velocity dispersions we adopt the iterative method 
used in \cite{milvang08}, using a biweight scale estimator \citep{beers90}.
The rest-frame velocity dispersion obtained for the main structure  is  quite robust and reliable:
$\sigma_{\rm rf}=522^{+111}_{-111} {\rm km \, s^{-1}}$.
However, 
we note that although in this work 13 spectra were added, the error in
sigma did not appreciably decrease from the value found in \cite{milvang08} (see \S\ref{maincl}), which implies that the
structure is not relaxed.
The velocity dispersions for all other structures are very uncertain,
%, while in all other cases the values are very uncertain, 
hence not useful for the following analysis. 
In \tab\ref{tab:mainstructures} additional information for the Cl~1103.7$-$1245 cluster is given.
For the sake of completeness, we mention that we obtained 
$\sigma_{\rm rf}=517^{+204}_{-124}{\rm km  \, s^{-1}}$ for the secondary structure 1c, 
$\sigma_{\rm rf}=167^{+65}_{-34} {\rm km \, s^{-1}}$ for structure  1d, and
$\sigma_{\rm rf}=34^{+18}_{-18} {\rm km \, s^{-1}}$ for structure  1e, although we insist these are highly unreliable. 

\begin{table*}[!t]
\caption{ $R_{200}$, $M_\sigma$, and $M_{lens}$ for Cl~1103.7$-$1245.\label{tab:mainstructures}}
\begin{center}
\begin{tabular}{lcccc}
\hline
\hline 
\multirow{2}{*}{ name} &  $\sigma_{\rm rf}$&$ R_{200}$	& $ M_\sigma$	& $M_{lens}$ \\%& ${\bs M_X}$ \\
			&($\rm km \, s^{-1}$)&      ($\rm Mpc$) 	&($M_\odot$)	&($M_\odot$)	\\% &($M_\odot$)	\\
\hline
Cl~1103.7$-$1245 		&$522^{+111}_{-111}$	& 0.76	$\pm$ 0.16	&$(1.42\pm 0.99) \times 10^{14}$	&$(7.25\pm 3.48)\times 10^{14}$ \\%	&\\
\hline
\hline
\end{tabular}
\end{center}
\vspace*{0.1ex}
{\footnotesize {\it Notes} --  The virial radius $R_{200}$, defined as the radius delimiting a 
sphere with interior mean density 200 times the critical 
density of the Universe at that redshift, was computed as in \cite{poggianti06}, the masses $M_\sigma$ was computed following \cite{finn05}, and $M_{lens}$ is from \cite{clowe06}.  Errrors were computed with the propagation of errors.  }
\end{table*}

The left panel of  \fig\ref{fig:histoall} shows the spatial distribution of galaxies in the entire field. The structures identified are indicated with the different colors. Most of the galaxies of the main structure are concentrated around the BCG (see below), while galaxies of the other structures are very spread and it is not possible to define the position of the center. 

It is interesting to note that, even though the secondary structure has a very similar velocity dispersion value as the main cluster, it has $\times 3$ fewer members. It is also close to the main structure both in redshift and 
in physical distance  (their distance is $\sim 13\sigma$).
This probably indicates that  the secondary structure is not quite a cluster yet, but some smaller system possibly infalling onto the main cluster at later time.

Due to the small number of members, 
we are not able to identify any substructure in any of the structures.  

In this section, we have investigated the structures found in the field, we can now proceed 
characterizing in detail the galaxy populations of two of them: the structures 1 and 1c.
 
\subsection{Analysis of the galaxy properties and stellar populations}
\subsubsection{The Cl~1103.7$-$1245 cluster }\label{maincl}
In this subsection we carefully analyze the properties of Cl~1103.7$-$1245, the main structure identified in the field. 
Our spectroscopic catalogue of this cluster now consists of 22 galaxies. When 
\cite{milvang08}  observed it, they found only 9 
members with robust redshift measurements. They measured  $z_{cl}=0.9586$ and 
 $\sigma_{cl}=534^{+101}_{-120}\, \kms $.
Putting all data together, 
 we find $z_{cl}=0.9580$ and $\sigma_{cl}=522 \pm 111\, \kms$.  
 
 One of our new members was also tentatively assigned  as cluster member by the ERGS project \citep{douglas10}, 
  using a lower resolution grism.
  
%Besides the measurements we determined from the spectroscopy and the morphologies,
Other pieces of information for this field can be found % are available from catalogues already published.
%A first characterization of the cluster is presented 
in \cite{white05} and 
% photometric redshift  information 
%is presented in 
\cite{pello09}. % and %, 24 $\mu m$ detections are characterized in
% \cite{finn10}. 
Using all of these, we  characterize the galaxy population of both the cluster 
and the secondary structure (see \S\ref{sect:gro}).

 For one galaxy member, {EDCSNJ1103464$-$1248034}, 
the photometry was not reliable,  
due to the presence of a galaxy close to it that altered its observed photometry, hence values
of magnitudes, colors, and mass are not reliable. It will be disregarded in the following analysis.

\cite{white05} gave the position of the BCG candidate for each \E cluster. 
As already found by the previous EDisCS data, the galaxy proposed to be the BCG for this cluster actually %turns out to be a cluster member, and it 
is 
the most luminous  cluster member galaxy. % of the structure.  
This galaxy is {EDCSNJ1103434$-$1245341} and its position is shown in the left panel of  \fig\ref{fig:histoall}. 
 It appears that the BCG lies towards the edge of the
 distribution of members, arguing that this cluster is not relaxed and that
 there may be a significant systematic error in the velocity
 dispersion. Another explanation to the fact that there are no
 objects on the right hand side of the field could be that that
 region has been observed during the old run, which was unfavorable
 to z=0.96, hence it was hard to find cluster members.

 \begin{figure}[!t]
\centering
\includegraphics[scale=0.56]{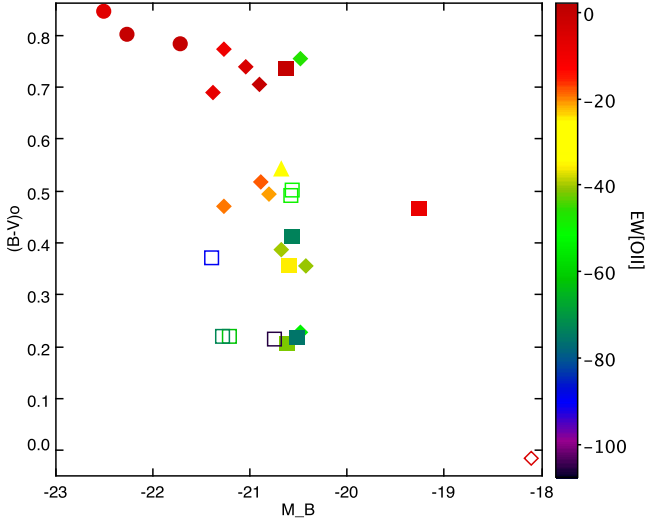}
\caption{Rest-frame $(B-V)_0$ vs.  $M_{B}$ diagram for spectroscopic members of the cluster 1 (filled symbols) and the secondary structure 1c (empty symbols), with also the morphology and  the [O{\sc ii}] (color coded) information. Circles: ellipticals. Diamonds: spirals. Squares: irregulars/peculiars. Triangles: galaxy out of the ACS image.
\label{fig:B_BV}}
\end{figure}
 In \fig\ref{fig:B_BV} the color magnitude diagram for  cluster members  (filled symbols) and  secondary structure members (empty symbols) is shown. Morphological types and  measured values of the EW([O{\sc ii}]) are also indicated, with different colors and symbols.
Rest-frame absolute $B$ magnitude and rest-frame $(B-V)_0$ are derived from photo-z fitting, fixing each galaxy 
redshift to be equal to the spectroscopic redshift of the cluster (\citealt{rudnick03, pello09}). 

Among the spectroscopic members,  three galaxies are ellipticals, eleven are spirals, 
seven are irregulars/peculiars
and one is located outside the ACS image.
It is interesting to note that  
no S0 galaxies are detected. 
All ellipticals have red colors, are located on the red sequence and do not have any significant star formation
([O{\sc ii}] was detected only in one elliptical galaxy with a very  
low $EW$, probably due to AGN activity). These are the brightest and the most massive galaxies
 (see below for the mass distribution). The BCG of the cluster is an elliptical galaxy. 
In contrast, late-type galaxies (spirals + irregulars/peculiars), which represent  the vast majority of the entire sample, % galaxies 
%($\sim 86\%$), %(spirals + irregulars + peculiars) 
cover a  wide color range. 
Six of them have quite red colors and are located on the red sequence.
This may be due to the presence of dust.
In fact, as proposed by \cite{wolf09}, optically-passive spirals in clusters, and dusty red galaxies 
appear to be basically
the same phenomenon. These objects are not truly
passive galaxies despite their red colors.
The other late-types belong to the blue cloud. All late-type galaxies show [O{\sc ii}] emission lines
in their spectrum, hence they are assumed to be  star-forming. %24$\mu m$ flux has been detected  for two of them.
On the whole, the [O{\sc ii}] in emission  has been measured for 17/22 galaxies. %, that is in 77.3\% of the total population.
All galaxies with no [O{\sc ii}] or with very weak emission lines are located on the red sequence.

\begin{figure}[!h]
\centering
\includegraphics[scale=0.4]{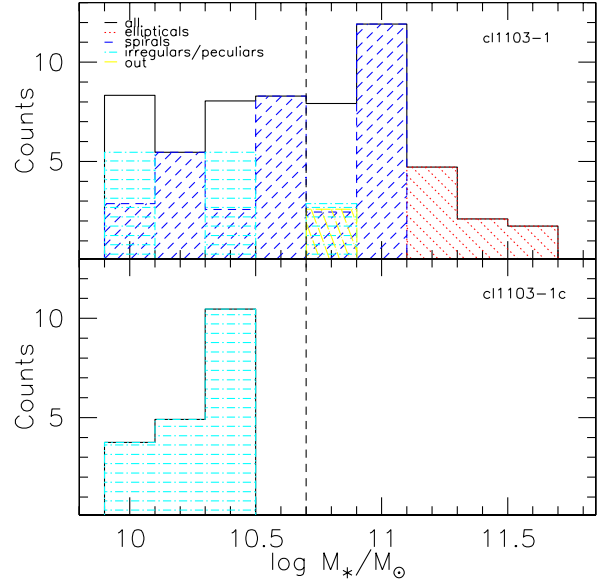}
 \caption{Weighted mass distribution of galaxies in the cluster. All (black line), elliptical (red line), spiral (blue line), and irregular/peculiar (cyan line) galaxies are shown. Also the galaxy without morphology is indicated (yellow line).  The mass completeness limit is indicated with a dashed vertical line. Upper panel:  cluster 1, lower panel: secondary structure 1c.
\label{fig:mass}}
\end{figure}

In the upper panel of \fig\ref{fig:mass} the mass distribution of cluster galaxies weighted for spectroscopic incompleteness
is shown. Galaxies cover a quite wide range of masses ($9.9\leq \log  M_{\ast}/M_{\odot} \leq 11.8$).
The mass distributions of galaxies of different morphological types are also indicated.
Ellipticals and late-types 
follow two distinct mass distributions. All ellipticals have $\log M_{\ast}/M_{\odot}\geq11.1$, while late-types have $\log M_{\ast}/M_{\odot}\leq 11.1$.

%We stress again that o
Our sample is complete in mass for all galaxy types down to $\log  M_{\ast}/M_{\odot} \sim 10.7$, and the differences in mass distribution between ellipticals and late-types are clearly noticeable above this limit. 

\subsubsection{The secondary structure Cl 1103.7-1245c} \label{sect:gro}
 In this subsection we study the galaxy population of Cl 1103.7-1245c, called secondary structure, which is represented by 
empty magenta pentagons in the left panel of \fig\ref{fig:histoall}. Our sample in 
this structure consists of seven galaxies. It is a diffuse structure and, except for three galaxies that are  close both in space and in redshift, 
members are  far from each other.  Due to the small number of galaxy members, the color magnitude diagram is  sparsely populated (see empty symbols in Fig. \ref{fig:B_BV}).   
The red sequence
 can not be detected; all galaxies are placed in the blue cloud. 

 One galaxy is located outside the ACS image and hence has no known morphology. All  others are 
irregulars/peculiars.
 Moreover, all galaxies in this structure have strong emission lines and are therefore  strongly star forming.  
 
 Comparing the results for the cluster and the secondary structure, we find that 
they consist of very different galaxy populations.
The cluster hosts both a population of passive galaxies and at least one  star-forming one, while in the secondary structure all galaxies are star-forming.
Also the mass distributions are different:  galaxies have systematically lower masses in the secondary structure than in the cluster.

To investigate whether the galaxy mass might be the main driver of the detected differences in the cluster and the secondary structure
we consider only cluster galaxies in the mass range of secondary structure galaxies ($9.9 \leq M_{\ast}/M_{\odot} \leq 10.5$).
These galaxies are all late-types and all of them are star-forming ([O{\sc ii}] detected). 
Four of them are spirals, 
four are irregulars/peculiars. In contrast, 
all secondary structure galaxies 
are irregulars and star forming. Thus, even if we consider only a common mass range, the galaxy population in cluster 1 and in structure 1c seem to be different, with less evolved galaxies in the latter. %In many ways, structure D galaxies seem to be more representative of field galaxies than of group/cluster ones. 

 Even though our sample is affected by low number statistic, it is worth noting that, neither in the cluster nor in the secondary structure do we detect galaxies in a transient phase, such as those moving from the blue cloud to the red sequence, 
or post-starburst galaxies (objects
characterized by strong Balmer lines in absorption and 
no emission lines, indicating that the star formation activity ended abruptly during the past $\sim$Gyr).  We note that the lack of E+A galaxies is consistent with
ÊÊ the number expected from lower redshift EDisCS clusters.

\subsection{Comparisons with lower redshift clusters}
The analysis presented in this paper allows us to significantly extend the redshift range of the EDisCS survey. 
It is therefore interesting to compare the galaxy population of  Cl~1103.7$-$1245 
with those in other \E clusters  to detect possible evolutionary trends.
For comparison we select all the \E structures that have a similar velocity dispersion but are located at lower redshifts.  
\tab\ref{tab:1} lists the clusters that satisfy these requirements.
\begin{table*}[!h]
\caption{EDisCS clusters used to compare galaxy populations. Number of members, magnitude limit and [O{\sc ii}] fractions for all clusters.\label{tab:1}}
\begin{center}
\begin{tabular}{lrl c c  cc}
\hline
\hline
name & z & ${\sigma}$ ($\kms$) &   $N$& $ M_V$ &  $f$([O{\sc ii}]) &$\langle\log (M_\ast/M_\odot)\rangle^1$ \\
\hline
Cl 1202.7$-$1224  & 0.4240  &  518$^{+92}_{-104    }$&  42.5  &$-20.5$&$0.27\pm 0.08$ & $10.97\pm 0.04$\\  
Cl 1138.2$-$1133a & 0.4548 &542$^{+63}_{-71     }$ & 48.8 &$-20.5$&$0.64\pm0.08$& $10.9\pm 0.05$\\
Cl 1059.2$-$1253  & 0.4564  &  510$^{+52}_{-56     }$  & 56.9&$-20.5$&$0.62 \pm 0.07$& $11.08\pm 0.02$\\ 
Cl 1018.8$-$1211  & 0.4734  &  486$^{+59}_{-63     }$ & 43.0&$-20.6$&$0.37 \pm 0.09$& $10.97\pm 0.04$\\  
Cl 1227.9$-$1138 &  0.6357  &  574$^{+72}_{-75     }$ &24.7 &$-20.7$&$0.63 \pm 0.01$ & $11.02\pm 0.06$\\ 
Cl 1054.4$-$1146 &  0.6972  & 589$^{+78}_{-70     }$ &80.1 &$-20.8$&$0.62 \pm 0.06$& $11.08\pm 0.04$\\  
Cl 1054.7$-$1245 & 0.7498  &  504$^{+113}_{-65    }$&37.8&$-20.8$&$0.18\pm 0.08$	& $11.19\pm 0.02$\\ 
Cl~1103.7$-$1245 & 0.9580&$522^{+111}_{-111}$&31.4 &$-21.0$&$0.76\pm 0.12$	& $11.06\pm 0.03$\\
\hline
\hline
\end{tabular}
\end{center}
\vspace*{0.1ex}
{\footnotesize $^1$Mean masses are computed above the common mass limit $\log (M_\ast/M_\odot)\geq 10.7$}.
\end{table*}

%\begin{small}
%\begin{landscape}
\begin{table*}[!t]
\caption{Morphological fractions (in \%) for all clusters, above the absolute  magnitude limit $M_V\leq -21$ at $z=0.95$, corrected for passive evolution. \label{tab:2}}
\begin{center}
\begin{tabular} {lrrrrrr}
\hline
\hline
 name	&	  compacts& ellipticals& S0s& spirals & irr/pec\\
\hline
Cl 1138.2$-$1133a &0$\pm$2 	&27$\pm$7	&11$\pm$6	&62$\pm$8	&0$\pm$2 	 \\	
Cl 1227.9$-$1138 &0$\pm$4	&13$\pm$8	&30$\pm$12	&58$\pm$13	& 0$\pm$4	\\
Cl 1054.4$-$1146&6$\pm$3	&32$\pm$6	&0$\pm$1	&53$\pm$6	&9$\pm$4	\\
Cl 1054.7$-$1245 &0$\pm$2	&43$\pm$5	&24$\pm$4	&33$\pm$5	&0$\pm$2		\\
 Cl~1103.7$-$1245 &0$\pm$3	&27$\pm$9	&0$\pm$3	&73$\pm$9	&0$\pm$3		 \\
\hline
\hline
\end{tabular}
\end{center}
\end{table*}

We first  adopt a common absolute magnitude limit (corrected for passive evolution\footnote{The passive evolution has been computed assuming
 the formation at z=5 and solar metallicity.}).
For Cl~1103.7$-$1245, the completeness magnitude limit is  fixed at $M_V=-21$. Hence, for the other clusters
we adopt a  limit which for passively-evolving galaxies corresponds to $M_V=-21$ at $z\sim 0.95$ (see \tab\ref{tab:1}). 
At any redshift, we consider only galaxies within $0.8R_{200}$. This corresponds to the spectroscopic coverage of the clusters at $z\sim0.4$. 
 
In this section we compare the completeness-corrected number of cluster members, the completeness-weighted fraction of galaxies with [O{\sc ii}] in emission,
 and, when available, the completeness-weighted morphological fractions.
In \tab\ref{tab:1} numbers and fractions for each cluster are given.

\begin{figure}[!b]
\centering
\includegraphics[scale=0.3,angle=-90]{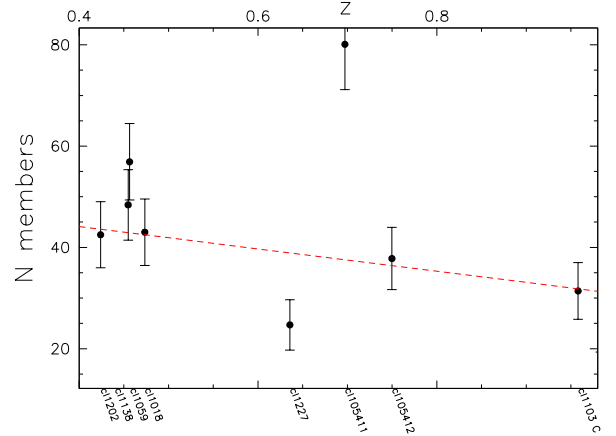}
\caption{Completeness-corrected number of cluster members as a function of redshift for EDisCS clusters with 
$\sigma=500 \, {\rm km\, s}^{-1}$. The predicted number of cluster members for the adopted magnitude limit is shown as the red dashed line. See text for details.\label{fig:cl_n}}
\end{figure}

In \fig\ref{fig:cl_n} the completeness-corrected number of members of each cluster is plotted as a function of $z$.
The number of galaxy members in clusters with similar velocity dispersion seems to decrease with 
redshift. The only exception in this trend is for Cl 1054.7-1245 that 
has many more members. This can be due to the presence of a secondary structure close to the main cluster  that is difficult to disentangle.
 
As found in \cite{poggianti10} in simulations, the number of cluster members per unit of cluster mass is constant both with redshift and cluster mass 
($N_{\rm memb}=\alpha M_{\rm sys}$).
These authors computed $\alpha = 20 \, {\rm galaxies}/10^{14} \, M_\odot$ for a magnitude limit $M_V=-20$.
At $z=0.45$, for a cluster $\sigma=500 \, \kms$, we estimate the cluster halo  mass to be $M_{\rm sys}=1.686 \times 10^{14} \, M_{\odot}$.\footnote{We compute  the mass of the system as in  \cite{poggianti06}.}
Considering Cl~1202.7$-$1224 and Cl~1018.8$-$1211 (whose mean redshift is $\langle z\rangle =0.45$ and mean velocity dispersion  is $\langle \sigma \rangle \sim 500 \, \kms$), we measure a number of members equal to 43 
for our magnitude limit of $M_V=-20.5$. Therefore $\alpha=25.5 \, {\rm galaxies}/ 10^{14} \,M_\odot$.  
As a consequence, the predicted number of galaxies for a cluster of the same velocity dispersion, but located at $z=0.95$ is $\sim 32$ (this calculation reflects the growth of the halo mass with cosmic time).  This is in good agreement with the observations shown in \fig\ref{fig:cl_n}: Cl~1103.7$-$1245 has 31.4 galaxies and a velocity dispersion  of $522 \pm 111 \, \kms$.

We now explore the evolution of the star-formation activity of the galaxies in these clusters using the [O{\sc ii}] emission line as a tracer. Looking at the fraction of galaxies with [O{\sc ii}] emission (see \tab\ref{tab:1}),\footnote{In computing the [O{\sc ii}] fractions, we consider only galaxies for which the [O{\sc ii}] line could be measured (even if absent).
However, in each cluster at most 1 or 2 galaxies had no EW([O{\sc ii}]) measurement.}
a slight trend with $z$ is detected: Cl~1103.7$-$1245 has a marginally higher fraction of star-forming galaxies than lower-$z$ clusters. %However, within the observational uncertainties, this cluster's star-formation activity does not seem to be abnormally high for its velocity dispersion (see \fig 4 of \citealt{poggianti06}).

Concerning galaxy morphologies  (see \tab\ref{tab:2}), Cl~1103.7$-$1245 seems to follow the trends 
found in \cite{desai07}: there is no significant redshift evolution in the elliptical galaxy fraction, 
but there is a clear decline in the S0 fraction with look-back-time.  Indeed, none of the spectroscopic members of Cl~1103.7$-$1245 is an S0, while we find a relatively-high fraction of late-type galaxies. However, the absence of S0s could simply be due to the small
sample size: at these redshifts the fraction of S0s found in clusters is quite small \citep{postman05}.
 
To compare galaxy stellar  masses between clusters we adopt mass-limited samples with $\log M_{\ast}/M_{\odot}\geq 10.7$, which are complete at all redshifts. The mean stellar masses of cluster galaxies are very similar for all the clusters
(\tab\ref{tab:1}). This suggests that the high-mass end of the galaxy population in clusters of similar velocity dispersion has not changed much in the redshift range explored, i.e., the most massive galaxies were already in place in $z\sim1$ clusters.

\section{Summary}
In this paper we present new spectroscopic observations 
for Cl~1103.7$-$1245, the most distant cluster of the EDisCS project ($z=0.9580$), 
complementing the previous analysis of \cite{milvang08}, whose observations targetted z=0.70 
and thus were somewhat biased against
galaxies in this cluster at z=0.96

%whose observations were biased
%towards this cluster.

From the spectra we measured the galaxies' redshifts,  [O{\sc ii}]$_{\lambda3727}$ and H$_{\delta}$ equivalent widths,
and spectroscopic completeness weights. These new data were complemented with 
information from \cite{white05} and \cite{pello09}, % and %, 24 $\mu m$ detections are characterized in
% \cite{finn10}, 
to characterise the galaxies' physical properties such as their stellar masses, star-formation activities and morphologies. This allowed us to
study the galaxy populations in the Cl~1103.7$-$1245 cluster ($z=0.9580$) and another structure at a similar redshift 
(Cl~1103.7$-$1245c at $z=0.9830$), and compared them with those of lower redshift clusters. The main results can be summarised as follows:
\begin{itemize}

\item 
The main cluster Cl~1103.7$-$1245 (for which we have 22 spectroscopic members) 
consists of two well separated populations: elliptical galaxies
located on the red sequence without any significant star-formation activity,
and active late-type galaxies,   that can be 
either  as red as the red sequence galaxies  or
cover  a  wide range of colors. 
All morphologically late-type galaxies  are star-forming. 
The cluster galaxies cover quite a broad range of masses ($9.9\leq \log  M_{\ast}/M_{\odot} \leq 11.8$). 
All ellipticals are very massive ($\log M_{\ast}/M_{\odot}\geq11.1$), while the late-types (spirals and irregulars) 
have $\log M_{\ast}/M_{\odot}\leq 11.1$.

\item Cl 1103.7-1245c, the secondary structure, (for which we have 7 spectroscopic members)
contains only irregular/peculiar galaxies, with some possibly undergoing  mergers. 
These galaxies are all star forming and located in the blue cloud. Their stellar masses are in
the $9.9 \leq \log M_{\ast}/M_{\odot} \leq 10.5$ range, systematically less massive than the galaxies in the main cluster.
\end{itemize}
We conclude that the main cluster and the secondary structure  
consist of very different galaxy populations, with clearly different 
star formation properties, morphologies and masses.

We have also compared the properties of the Cl~1103.7$-$1245 cluster galaxies with those of 
the galaxy populations in EDisCS clusters at lower redshift and comparable velocity dispersions ($\sigma_{\rm clus}\simeq500 \, \kms$).
Considering absolute-magnitude-limited samples ($M_V \leq -21$ at $z\sim 0.95$), and galaxies within $0.8 R_{200}$, the main results are: 
\begin{itemize}
\item 
The number of member galaxies in clusters with similar velocity dispersions decreases slightly with 
redshift. This is a consequence of the fact that the cluster mass is expected to grow with time and 
the prediction from  simulations that the number of galaxies per unit of cluster mass 
should be independent of redshift and cluster halo mass \citep{poggianti10}. 
\item 
The morphologcal mix and the star-formation activity in the cluster galaxies follows the trends found at lower redshifts.
The fraction of elliptical galaxies remains approximately constant with redshift, while the fraction of S0s declines as the fraction of later types increases. We found no spectroscopically-confirmed S0s in Cl~1103.7$-$1245. Furthermore, the fraction of star-forming galaxies in this cluster is marginally higher than in lower-redshift clusters of similar $\sigma_{\rm clus}$. 
\end{itemize}

Finally,  comparing mass-limited galaxy samples with $ \log M_{\ast}/M_{\odot} \leq 10.7$, we found that  
the galaxies in clusters of similar $\sigma_{\rm clus}$ have similar average stellar masses, 
suggesting  that the most massive galaxies were already in place in at $z\sim 1$ clusters.

\begin{acknowledgements}
We thank the referee for her/his useful comments which helped us to
improve our manuscript.
BV and BMP acknowledge financial support from ASI contract I/016/07/0 
and ASI-INAF I/009/10/0. BV also acknowledges financial support from the
Fondazione Ing. Aldo Gini and thanks the School of Physics and Astronomy, Univeristy of Nottingham, 
for a very pleasant and productive stay
during which part of the work presented in this paper was carried
out. The Dark Cosmology center is funded by the Danish National Research Foundation.
BMJ acknowledges support from the ERC-StG grant EGGS-278202.

\end{acknowledgements}

\bibliographystyle{aa}
 \bibliography{biblio_spec}
\end{document}